\documentclass{ifacconf}

\usepackage{graphicx}      
\usepackage{natbib}        
\usepackage{amsfonts, amsmath, amssymb, subfigure}
\usepackage{color, tikz, algorithm, algpseudocode}
\usetikzlibrary{shapes.geometric, arrows}

\newtheorem{assumption}{Assumption}

\theoremstyle{definition}
\newtheorem{definition}{Definition}
\theoremstyle{remark}
\newtheorem{remark}{Remark}

\tikzstyle{startstop} = [rectangle, rounded corners, minimum width=1cm, minimum height=1cm,text centered, draw=black, fill=red!30]
\tikzstyle{io} = [rectangle, minimum width=1.5cm, minimum height=1cm, text centered, draw=black, fill=green!10]
\tikzstyle{process} = [rectangle, minimum width=1.5cm, minimum height=1cm, text width=3.5cm, text centered, draw=black, fill=orange!30]
\tikzstyle{summer} = [circle, minimum width=0.75cm, minimum height=0.75cm, draw=black, fill=green!1]
\tikzstyle{route} = [circle, fill, inner sep=1.5pt]
\tikzstyle{arrow} = [thick,->,>=stealth]

\begin{document}
	\begin{frontmatter}
		
		\title{Exponentially Stable Combined Adaptive Control under Finite Excitation Condition} 
		
		\thanks[footnoteinfo]{\copyright 2025 the authors. This work has been accepted to IFAC for publication under a Creative Commons Licence CC-BY-NC-ND.}
		
		\author[First]{Manish Patel} 
		\author[Second]{Arnab Maity}
		
		\address[First]{Indian Institute of Technology Bombay, Mumbai, India
			(e-mail: pmanish@aero.iitb.ac.in).}
		\address[Second]{Indian Institute of Technology Bombay, Mumbai, India
			(e-mail: arnab@aero.iitb.ac.in)}
		
		\begin{abstract}                
			The parameter convergence relies on a stringent persistent excitation (PE) condition in adaptive control. Several works have proposed a memory term in the last decade to translate the PE condition to a feasible finite excitation (FE) condition. This work proposes a combined model reference adaptive control for a class of uncertain nonlinear systems with an unknown control effectiveness vector. The closed-loop system is exponentially stable under the FE condition. The exponential rate of convergence is independent of the excitation level of the regressor vector and is lower-bounded in terms of the system parameters and user-designed gains. Numerical simulation is illustrated, validating the results obtained with the proposed adaptive control.
		\end{abstract}
		
		
	\end{frontmatter}
	
	\section{Introduction}
	The design of an adaptive controller for solving regulation and tracking problems is well-established for systems with uncertain parameters \citep{ioannou1996robust}. The primary objective is to achieve zero tracking/regulation error by designing a control law driven by the estimation of uncertain parameters. Converging parameter estimates to their ideal values allows a unique control law design and provides robustness to the non-parametric bounded uncertainties. However, persistent excitation (PE) is the necessary and sufficient condition for parameter convergence \citep{narendra2012stable}. The PE condition relies on the future closed-loop signals; hence, it is an infeasible condition to satisfy in practice. The PE condition is translated to a class of reference trajectories \citep{boyd1986necessary}; however, the reference trajectories depend on the desired task, and enriching the transient profile of the reference trajectories \citep{gallegos2024relaxed} may degrade the closed-loop system performance. \par
	
	Several works have proposed a memory term \citep{ortega2020modified} in the gradient/least-squares-based adaptation law for estimating uncertain parameters. The memory term accumulates the past closed-loop signals and supplies a symmetric positive definite coefficient matrix\footnote{The gradient/least-squares-based adaptation law for estimating uncertain parameters produces a first-order ordinary differential equation representing the parameter estimation error dynamics (PEE).} to parameter estimation error dynamics (PEE) under a finite excitation (FE) condition \citep{chowdhary2013concurrent}, \citep{pan2018composite}. The FE condition depends only on past closed-loop signals that can be checked online, making it feasible. The coefficient matrix is symmetric positive definite, which may cause undesirable $\left(5.3, \, \text{\citep{ortega2020modified}} \right)$ high-gain adaptation in the following ways. First, the coefficient matrix contains closed-loop signals as its entries. The closed-loop signals vary with the reference trajectories and the initial conditions. The choice of adaptation gains for a set of reference trajectories and initial conditions may result in high-gain adaptation for another set, and it will surely happen if the regressor vector satisfies the PE condition. Second, choosing adaptation gains for fast parameter convergence of an individual parameter estimate may cause high-gain adaptation of another parameter estimate as the coefficient matrix is coupled. \par
	
	Creating the memory term using dynamic regressor extension and mixing (DREM) \citep{aranovskiy2016parameters}, partially solves the above-stated issues. It supplies a positive definite diagonal coefficient matrix, therefore, fast individual parameter convergence is possible. However, the diagonal entries depend on the closed-loop signals. Thus, the first route of the high-gain adaptation discussed above is unresolved with the DREM method. Time-varying adaptation gains \citep{glushchenko2022exponentially}, \citep{patel2022parameter} may serve the purpose but the gains may assume unwanted high values causing instability mechanisms. Further, in creating the memory term, the DREM method requires several filters proportional to the unknown system parameters, and the selection of the filters' frequency is not systematic and is more involved. \par
	
	The above-discussed methods to relax the PE condition to the FE condition are illustrated to solve regressor equations \citep{marino2022exponentially}, \citep{ortega2022new}, adaptive control with a known control effectiveness vector \citep{cho2017composite}, and combined adaptive control with a known lower bound of the control effectiveness vector \citep{roy2017combined}, \citep{gerasimov2018relaxing}. In combined adaptive control, the adaptation laws for the controller gains estimate utilize the tracking error and the system parameter estimates \citep{duarte1989combined}. The appearance of the error term between the system parameter estimate and the ideal value in the $\dot{V}$ expression\footnote{The time-derivative of the Lyapunov function.}, restricts achieving exponential stable combined adaptive control; the closed-loop errors converge to the origin only exponentially fast. Further, robustness to bounded non-parametric uncertainties is complex to analyze for such design methods. To the best of the author's knowledge, the design of exponential stable combined adaptive control with known signs of the control effectiveness vector is unavailable. It is not because of its obviousness but it can be attributed to the hindrance in achieving the desired results. \par
	
	Motivated by the above-stated issues, we put an effort into proposing a solution that collectively resolves them. We propose an algorithm that extracts the system's ideal parameter values under the FE condition. These values are then fed to the adaptation laws driving the update of the controller gains estimate. Here, in our methodology, the adaptation laws are fed with the system parameter ideal values and not its estimate; hence, avoiding the appearance of the error term between the parameter estimates and ideal values in the $\dot{V}$ expression. This benefits achieving exponential stability, and the convergence rate depends only on the constant system parameters and user-designed gains. Therefore, the developed algorithm combined with the proposed methodology for the adaptation laws is beneficial three-fold, as described below.
	\begin{enumerate}
		\item The proposed algorithm creates a memory term that supplies an identity matrix to the PEE under the FE condition; hence, closing both possibilities of high-gain adaptation.
		\item The tracking error and the extracted ideal system parameters drive the adaptation laws; achieving an exponentially stable closed-loop system under the FE condition.
		\item The exponential decay rate is independent of the regressor vector's excitation level under the FE condition. It implies that if the regressor vector satisfies the FE condition for a set of reference trajectories and initial conditions; we achieve an exponential decay rate with a fixed lower bound.
	\end{enumerate}
	
	Next, we have put a brief section [\textit{Section} $\ref{section:Preliminaries}$] for the definitions, and notations, followed by the problem formulation [\textit{Section} $\ref{section:Problem_Formulation}$], and then the main results are discussed [\textit{Section} $\ref{section:Main_Results}$]. The obtained results with the proposed combined adaptive control are validated using numerical simulation in \textit{Section} $\ref{section:Simulation_Results}$, and \textit{Section} $\ref{section:Conclusion}$ concludes the article.
	
	\section{Preliminaries}
	\label{section:Preliminaries}
	The following notations and definitions are used throughout this article, a scalar variable is denoted in small  $q$, a vector in small boldface $\mathbf{q}$, and a matrix in capital boldface $\mathbf{Q}$. The Euclidean norm of a vector is denoted by $\|\mathbf{q}\|$, in addition, $\lambda_{\text{min}}(\mathbf{Q})$, $\lambda_{\text{max}}(\mathbf{Q})$, and $\det(\mathbf{Q})$, denote the minimum and maximum eigenvalues, and determinant of a square matrix $\mathbf{Q}$, respectively. An orthogonal matrix $\mathbf{Q}$ satisfies $\mathbf{Q} \mathbf{Q}^{T} = \mathbf{I}$, where $\mathbf{I}$ is an identity matrix.
	\begin{definition}[\citep{yuan1977probing}]
		A bounded signal $\boldsymbol{\phi}(t) \in \mathbb{R}^{n}$ is said to be persistently exciting if there exist positive constants $\gamma$ and $T$ such that for every $t > 0$, there is a sequence of $q$ numbers in $t_{i} \in [t,t+T]$, $i \in \mathbb{N}$, with
		\begin{equation}
			\left\| \begin{bmatrix} \boldsymbol{\phi}(t_{1}) & \boldsymbol{\phi}(t_{2}) & ... & \boldsymbol{\phi}(t_{q}) \end{bmatrix}^{-1} \right\| \leq \gamma.
			\label{eq:pe_definition}
		\end{equation}
	\end{definition}
	\begin{definition}
		A bounded signal $\boldsymbol{\phi}(t) \in \mathbb{R}^{n}$ is said to be finitely exciting if there exist positive constants $\gamma$ and $T$ so that there is at least a sequence of $q$ numbers in $t_{i} \in [t,t+T]$, $i \in \mathbb{N}$, with
		\begin{equation}
			\left\| \begin{bmatrix} \boldsymbol{\phi}(t_{1}) & \boldsymbol{\phi}(t_{2}) & ... & \boldsymbol{\phi}(t_{q}) \end{bmatrix}^{-1} \right\| \leq \gamma,
			\label{eq:fe_definition}
		\end{equation}
		where $\gamma$ is the excitation level of the signal $\boldsymbol{\phi} \left(t\right)$.
	\end{definition}
	
	\section{Problem Formulation}
	\label{section:Problem_Formulation}
	In this section, we define the system dynamics, the control objectives, and the assumptions taken. Consider a class of uncertain nonlinear single-input systems,
	\begin{equation}
		\dot{\mathbf{x}} = \mathbf{A} \mathbf{x} + \mathbf{b} k_{p} \left( u + \Delta \left( \mathbf{x} \right) \right),
		\label{eq:system_dynamics}
	\end{equation}
	where $\mathbf{x} \in \mathbb{R}^{n}$ and $u \in \mathbb{R}$ denote the state vector and the control input of the system, respectively. The state matrix $\mathbf{A} \in \mathbb{R}^{n \times n}$ is unknown, the control vector $\mathbf{b}k_{p} \in \mathbb{R}^{n}$ contains known vector $\mathbf{b} \in \mathbb{R}^{n}$, and unknown constant $k_{p} \in \mathbb{R}$, however, sign of $k_{p}$ is known. The nonlinear uncertain term $\Delta \left( \mathbf{x} \right) \in \mathbb{R}$ is a matched uncertainty. 
	\begin{assumption}
		The nonlinear matched uncertain term $\Delta \left( \mathbf{x} \right)$, is linearly parameterizable as $\Delta \left( \mathbf{x} \right) = \boldsymbol{\theta}^{T} \boldsymbol{\phi} \left( \mathbf{x} \right)$, where $\boldsymbol{\phi} \left( \mathbf{x} \right): \mathbb{R}^{n} \to \mathbb{R}^{p}$ is a known mapping, and called the regressor vector, and $\boldsymbol{\theta} \in \mathbb{R}^{p}$ contains unknown constants.
	\end{assumption}
	\begin{assumption}
		The given system $\eqref{eq:system_dynamics}$ is controllable that is, $\left( \mathbf{A}, \mathbf{b} k_{p} \right)$ is a controllable pair hence $\mathbf{b}^{T} \mathbf{b} > 0$. The sign of the unknown constant $k_{p}$ is known and is denoted by $k_{p}'$.
	\end{assumption}
	
	The primary control objective is to design a control law $u$, such that the state vector $\mathbf{x}$ of the system tracks a reference state trajectory $\mathbf{x}_{r}$, generated online from the following reference dynamics,
	\begin{equation}
		\dot{\mathbf{x}}_{r} = \mathbf{A}_{r} \mathbf{x}_{r} + \mathbf{b}_{r} r,
		\label{eq:reference_dynamics}
	\end{equation}
	where $\mathbf{A}_{r} \in \mathbb{R}^{n \times n}$ is a Hurwitz matrix, $\mathbf{x}_{r} \in \mathbf{R}^{n}$ is the reference state trajectory, and $r \in \mathbb{R}$ denotes a bounded, piece-wise continuous reference signal. $\mathbf{A}_{r}$ is a Hurwitz matrix hence satisfies,
	\begin{equation}
		\mathbf{A}_{r}^{T} \mathbf{P} + \mathbf{P} \mathbf{A}_{r} + \mathbf{Q} = \mathbf{0},
		\label{eq:lyapunov_equation}
	\end{equation}
	where $\mathbf{P}$ and $\mathbf{Q}$ are positive definite matrices. The well known control law with feedback term $\mathbf{k}_{x} \in \mathbb{R}^{n}$ and feedforward term $k_{r} \in \mathbb{R}$ is,
	\begin{equation}
		u = \mathbf{k}_{x}^{T} \mathbf{x} + k_{r} r - \boldsymbol{\theta}^{T} \boldsymbol{\phi} \left( \mathbf{x} \right).
		\label{eq:control_law_ideal}
	\end{equation}
	Let the tracking error denoted by $\mathbf{e}$ is defined as $\mathbf{e}=\mathbf{x}-\mathbf{x}_{r}$, the tracking error dynamics using $\eqref{eq:system_dynamics}$, $\eqref{eq:reference_dynamics}$ and $\eqref{eq:control_law_ideal}$,
	\begin{equation}
		\dot{\mathbf{e}} = \mathbf{A}_{r} \mathbf{e} + \left( \mathbf{A} + \mathbf{b} k_{p} \mathbf{k}_{x}^{T} - \mathbf{A}_{r} \right) \mathbf{x} + \left( \mathbf{b} k_{p} k_{r} - \mathbf{b}_{r} \right) r.
		\label{eq:error_dynamics_ideal}
	\end{equation}
	The origin of the tracking error is exponentially stable when the controller gains $\mathbf{k}_{x}$ and $k_{r}$ satisfy the following matching condition,
	\begin{equation}
		\mathbf{A} + \mathbf{b} k_{p} \mathbf{k}_{x}^{T} = \mathbf{A}_{r}, \quad \mathbf{b} k_{p} k_{r} = \mathbf{b}_{r}.
		\label{eq:matching_condition}
	\end{equation}
	However, $\mathbf{A}$ and $k_{p}$ are unknown, the matching condition $\eqref{eq:matching_condition}$ would not generate the controller gains, and $\boldsymbol{\theta}$ is also unknown, hence the control law $\eqref{eq:control_law}$ is modified as below,
	\begin{equation}
		u = \hat{\mathbf{k}}_{x}^{T} \mathbf{x} + \hat{k}_{r} r - \hat{\boldsymbol{\theta}}^{T} \boldsymbol{\phi} \left( \mathbf{x} \right),
		\label{eq:control_law}
	\end{equation}
	where $\hat{\mathbf{k}}_{x}$, $\hat{k}_{r}$, and $\hat{\boldsymbol{\theta}}$ denote the estimate of $\mathbf{k}_{x}$, $k_{r}$, and $\boldsymbol{\theta}$, respectively. Let the error in the estimation of the controller gains denoted by $\tilde{\mathbf{k}}_{x}$, $\tilde{k}_{r}$, and $\tilde{\boldsymbol{\theta}}$ are defined as $\tilde{\mathbf{k}}_{x} = \hat{\mathbf{k}}_{x} - \mathbf{k}_{x}$, $\tilde{k}_{r} = \hat{k}_{r} - k_{r}$, and $\tilde{\boldsymbol{\theta}} = \hat{\boldsymbol{\theta}} - \boldsymbol{\theta}$, respectively. The tracking error dynamics $\dot{\mathbf{e}}$ in $\eqref{eq:error_dynamics_ideal}$ is as follows after putting $\eqref{eq:control_law}$ in $\eqref{eq:system_dynamics}$, and using $\eqref{eq:reference_dynamics}$ and $\eqref{eq:matching_condition}$,
	\begin{equation}
		\dot{\mathbf{e}} = \mathbf{A}_{r} \mathbf{e} + \mathbf{b} k_{p} \tilde{\mathbf{k}}_{x}^{T} \mathbf{x} + \mathbf{b} k_{p} \tilde{k}_{r} r - \mathbf{b} k_{p} \tilde{\boldsymbol{\theta}}^{T} \boldsymbol{\phi}.
		\label{eq:tracking_error_dynamics}
	\end{equation}
	The second control objective is the convergence of the controller gains' estimate $\hat{\mathbf{k}}_{x}$, $\hat{k}_{r}$, and $\hat{\boldsymbol{\theta}}$ to their ideal values $\mathbf{k}_{x}$, $k_{r}$, and $\boldsymbol{\theta}$, respectively, under the finite excitation condition of the regressor vector,
	\begin{equation}
		\boldsymbol{\varphi} = \begin{bmatrix} \mathbf{x}^{T} & u & \boldsymbol{\phi}^{T} \left( \mathbf{x} \right) \end{bmatrix}^{T} \in \mathbb{R}^{n+1+p}.
		\label{eq:regressor_vector}
	\end{equation}
	\begin{assumption}
		The regressor vector $\boldsymbol{\varphi}$ satisfies the finite excitation condition. Therefore from $\eqref{eq:fe_definition}$, there exists positive constants $\gamma$ and $T$ so that there is at least a sequence of $q$ numbers in $t_{i} \in \left[ t, t+T \right]$, $i \in \mathbb{N}$, satisfying the following inequality,
		\begin{equation}
			\left\| \begin{bmatrix} \boldsymbol{\varphi} \left( \mathbf{x} \left(t_{1}\right), t_{1}  \right) & \hdots & \boldsymbol{\varphi} \left( \mathbf{x} \left(t_{q}\right), t_{q}  \right) \end{bmatrix}^{-1} \right\| \leq \gamma,
			\label{eq:regressor_fe}
		\end{equation}
		where $q \geq n+1+p$, and $\gamma$ is the excitation level of the regressor vector $\boldsymbol{\varphi}$.
	\end{assumption}
	Next, follows the main results of the work where we propose an algorithm for extracting the system parameters' ideal values, and the adaptation laws are designed to fulfill the control objectives under the stated assumptions.
	
	\section{Main Results}
	\label{section:Main_Results}
	The gradient-based adaptation law \citep{ioannou1996robust} that updates the controller gains' estimate in the direction of minimizing the tracking error is,
	\begin{equation}
		\begin{bmatrix} \dot{\hat{\mathbf{k}}}_{x}^{T} & \dot{\hat{k}}_{r} & \dot{\hat{\boldsymbol{\theta}}}^{T}
		\end{bmatrix}^{T} = \begin{bmatrix} - \mathbf{x}^{T} & -r & \boldsymbol{\phi}^{T} \end{bmatrix}^{T} \mathbf{e}^{T} \mathbf{P} \mathbf{b} k_{p}',
		\label{eq:adaptation_law_gradient}
	\end{equation}
	generating following error dynamics,
	\begin{equation}
		\begin{bmatrix} \dot{\tilde{\mathbf{k}}}_{x}^{T} & \dot{\tilde{k}}_{r} & \dot{\tilde{\boldsymbol{\theta}}}^{T}
		\end{bmatrix}^{T} = \begin{bmatrix} - \mathbf{x}^{T} & -r & \boldsymbol{\phi}^{T} \end{bmatrix}^{T} \mathbf{e}^{T} \mathbf{P} \mathbf{b} k_{p}'.
		\label{eq:estimation_error_dynamics_gradient}
	\end{equation}
	
	Next, we briefly state the standard result in the adaptive control for a smooth transition from the existing results to the ones achieved by the proposed adaptive controller.
	
	\begin{prop}
		Consider the class of uncertain nonlinear system $\eqref{eq:system_dynamics}$, driven by the control law $\eqref{eq:control_law}$, and the estimate of the controller gains are updated by the adaptation laws $\eqref{eq:adaptation_law_gradient}$. The origin of the closed-loop system described by $\eqref{eq:tracking_error_dynamics}$ and $\eqref{eq:estimation_error_dynamics_gradient}$ is uniformly stable and that of tracking error is asymptotically stable.
		\label{prop:proposition_gradient}
	\end{prop}
	
	\begin{pf}
		Let the Lyapunov function in terms of the tracking and estimation errors be,
		\begin{equation}
			V = \mathbf{e}^{T} \mathbf{P} \mathbf{e} + \left| k_{p} \right| \tilde{\mathbf{k}}_{x}^{T} \tilde{\mathbf{k}}_{x} + \left| k_{p} \right| \tilde{k}_{r}^{2} + \left| k_{p} \right| \tilde{\boldsymbol{\theta}}^{T} \tilde{\boldsymbol{\theta}},
			\label{eq:lyapunov_function}
		\end{equation}
		where $\left| k_{p} \right|$ denotes the absolute value of $k_{p}$, the time derivative of the Lyapunov function along the error trajectories $\eqref{eq:tracking_error_dynamics}$ and $\eqref{eq:estimation_error_dynamics_gradient}$, and using $\eqref{eq:lyapunov_equation}$, we obtain
		\begin{equation}
			\dot{V} \leq -\lambda_{\text{min}} \left( \mathbf{Q} \right) \left\| \mathbf{e} \right\|^{2}.
		\end{equation}
		Let the combined error vector, denoted by $\boldsymbol{\chi}$, is defined as
		\begin{equation}
			\boldsymbol{\chi} = \begin{bmatrix} \mathbf{e}^{T} & \tilde{\mathbf{k}}_{x}^{T} & \tilde{k}_{r} & \tilde{\boldsymbol{\theta}}^{T} \end{bmatrix}^{T},
			\label{eq:combined_error}
		\end{equation}
		the Lyapunov function satisfies, $\text{min} \left\{ \lambda_{\text{min}} \left( \mathbf{P} \right), \left| k_{p} \right| \right\} \left\| \boldsymbol{\chi} \right\|^{2}$ $\leq V \leq$ $\text{max} \left\{ \lambda_{\text{max}} \left( \mathbf{P} \right), \left| k_{p} \right| \right\} \left\| \boldsymbol{\chi} \right\|^{2}$, hence $V$ is \textit{positive definite} and \textit{decrescent} \citep{slotine1991applied}, also $\dot{V} \leq 0$, that is, $\dot{V}$ is \textit{negative semi-definite}, therefore using [Theorem $4.1$, \citep{slotine1991applied}], the origin of the closed-loop system is uniformly stable. Further using \textit{Barbalat's Lemma}, the origin of the tracking error is asymptotically stable \citep{slotine1991applied}, this completes the proof. \hfill $\blacksquare$ 
	\end{pf}
	In combined adaptive control, the adaptation laws are driven by the tracking error and the system parameters' ideal values. Since the ideal values are unknown, we propose an algorithm that extracts the ideal values under the FE condition.
	
	\subsection{Extraction of System Parameters}
	The system dynamics $\eqref{eq:system_dynamics}$ can be written in linear-in-parameter form,
	\begin{equation}
		\dot{\mathbf{x}} = \underbrace{\begin{bmatrix} \mathbf{A} & \mathbf{b} k_{p} & \mathbf{b} k_{p} \boldsymbol{\theta}^{T} \end{bmatrix}}_{\mathbf{W}^{T}} \boldsymbol{\varphi},
		\label{eq:system_lip}
	\end{equation}
	where $\mathbf{W} \in \mathbb{R}^{q \times n}$, $q=n+1+p$, and $\boldsymbol{\varphi} \in \mathbb{R}^{q}$ is defined in $\eqref{eq:regressor_vector}$. The regressor vector $\boldsymbol{\varphi}$ is measurable however $\dot{\mathbf{x}}$ is not, hence both the signals are passed through a stable filter $\left(f/\left(s+f\right),\, s:=d/dt\right)$. Let $\mathbf{y}_{f}$ and $\boldsymbol{\varphi}_{f}$ denote the filtered signals of $\dot{\mathbf{x}}$ and $\boldsymbol{\varphi}$, respectively, and their dynamics are as follows,
	\begin{equation}
		\begin{aligned}
			\dot{\mathbf{x}}_{f} =& -f \mathbf{x}_{f} + f \mathbf{x},\\
			\mathbf{y}_{f} =& f \mathbf{x} - e^{-f\left(t-t_{0}\right)} f \mathbf{x} \left(t_{0}\right) - f \mathbf{x}_{f},\\
			\dot{\boldsymbol{\varphi}}_{f} =& -f \boldsymbol{\varphi}_{f} + f \boldsymbol{\varphi},
		\end{aligned}
		\label{eq:filtered_dynamics_detailed}
	\end{equation}
	where $f \in \mathbb{R}^{+}$ is the cut-off frequency of the filter and initial condition of the signals $\mathbf{x}_{f}$ and $\boldsymbol{\varphi}_{f}$ are considered to be zero; hence, we can write
	\begin{equation}
		\mathbf{y}_{f} = \mathbf{W}^{T} \boldsymbol{\varphi}_{f}.
		\label{eq:filtered_dynamics}
	\end{equation}
	Let the regressor signal $\boldsymbol{\varphi}$ satisfies FE condition, [\textit{Lemma} 6.8, \citep{narendra2012stable}] infers $\boldsymbol{\varphi}_{f}$ also satisfies FE condition, then there exists a full rank matrix $\eqref{eq:regressor_fe}$, that is converted to an orthogonal matrix $\mathbf{\Phi}_{b} \in \mathbb{R}^{q \times q}$ using \textit{Modified Gram-Schmidt (MGS)} \citep{johnson1985matrix}, and a corresponding transformation is done to obtain another matrix $\mathbf{Y}_{b} \in \mathbb{R}^{n \times q}$.
	Let the matrices $\mathbf{Y}_{b}$ and $\mathbf{\Phi}_{b}$ are
	\begin{equation}
		\mathbf{Y}_{b} = \begin{bmatrix} \mathbf{Y}_{b_{1}} & \hdots & \mathbf{Y}_{b_{q}} \end{bmatrix}, \quad \mathbf{\Phi}_{b} = \begin{bmatrix} \mathbf{\Phi}_{b_{1}} & \hdots & \mathbf{\Phi}_{b_{q}} \end{bmatrix},
		\label{eq:mgs_stacked}
	\end{equation}
	where $\mathbf{Y}_{b_{i}} \in \mathbb{R}^{n}$, $\mathbf{\Phi}_{b_{i}} \in \mathbb{R}^{q}$, $i=\left\{1,2,\hdots,q\right\}$, are computed using the \textit{Algorithm $\ref{alg:identification}$}, that uses the following recursive calculations,
	\begin{equation}
		\begin{aligned}
			\mathbf{v}_{l} =& \mathbf{v}_{l-1} - \mathbf{\Phi}_{b_{l-1}}^{T} \mathbf{v}_{l-1} \mathbf{\Phi}_{b_{l-1}}, \quad 	\mathbf{\Phi}_{b_{i}} = \dfrac{\mathbf{v}_{i}}{\left\|  \mathbf{v}_{i} \right\|}, \\
			\mathbf{y}_{l} =& \mathbf{y}_{l-1} - \mathbf{\Phi}_{b_{l-1}}^{T} \mathbf{v}_{l-1} \mathbf{Y}_{b_{l-1}}, \quad \mathbf{Y}_{b_{i}} = \dfrac{1}{\left\|  \mathbf{v}_{i} \right\|}  \mathbf{y}_{i},
		\end{aligned}
		\label{eq:memory_term_creation}
	\end{equation}
	where $l=\left\{1,2,\hdots,i\right\}$ for every $i$, $\mathbf{v}_{0} = \boldsymbol{\varphi}_{f}$, $\mathbf{y}_{0} = \mathbf{y}_{f}$, $\mathbf{\Phi}_{b_{0}} = 0$, $\mathbf{Y}_{b_{0}}=0$, and $i=\left\{1,2,\hdots,q\right\}$.  \par
	
	The filtered signals $\mathbf{y}_{f}$ and $\boldsymbol{\varphi}_{f}$ are stored into the matrices $\mathbf{Y}_{b}$ and $\mathbf{\Phi}_{b}$ using \textit{Algorithm} $\ref{alg:identification}$, so that following holds,
	\begin{equation}
		\mathbf{Y}_{b} = \mathbf{W}^{T} \mathbf{\Phi}_{b},
			\label{eq:filtered_dynamics_basis}
	\end{equation}
	where $\mathbf{\Phi}_{b}$ is an orthogonal matrix.
	
	
	Next, post-multiplying $\eqref{eq:filtered_dynamics_basis}$ by $\mathbf{\Phi}_{b}^{T}$, we obtain
	\begin{equation}
		\mathbf{Y}_{m} = \mathbf{W}^{T} \mathbf{\Phi}_{m},
		\label{eq:filtered_dynamics_memory}
	\end{equation}
	where $\mathbf{\Phi}_{m} = \mathbf{\Phi}_{b} \mathbf{\Phi}_{b}^{T}$ and $\mathbf{Y}_{m} = \mathbf{Y}_{b} \mathbf{\Phi}_{b}^{T}$. Using the property of the orthogonal matrices, $\mathbf{\Phi}_{b} \mathbf{\Phi}_{b}^{T} = \mathbf{I}$ hence $\mathbf{\Phi}_{m} = \mathbf{\Phi}_{b} \mathbf{\Phi}_{b}^{T} = \mathbf{I}$, inferring $\mathbf{Y}_{m} = \mathbf{W}^{T} \mathbf{\Phi}_{m} = \mathbf{W}^{T}$. Since, the finite excitation condition is assumed [\textit{Assumption} $3$] to achieve after a finite time interval $t=t_{q}>t_{0}$, then
			\begin{equation}
				\mathbf{\Phi}_{m} = \mathbf{I} \implies \mathbf{Y}_{m} = \mathbf{W}^{T}, \quad \forall t \geq t_{q}.
				\label{eq:system_parameter_extraction}
			\end{equation} 
	
	\begin{algorithm}
		\begin{algorithmic}[1]
			\State Initialize $\epsilon_{1}$ and $\epsilon_{2}$
			\State $q \gets$ Number of unknown parameters
			\State $i \gets 1$
			\If{i $\leq$ q}
			\If{$\|\boldsymbol{\varphi}_{f}\| > \epsilon_{1}$}
			\State $\mathbf{v}_{0}$ $\gets$ $\boldsymbol{\varphi}_{f}$
			\State Compute $\mathbf{v}_{i}$ recursively using $\eqref{eq:memory_term_creation}$ 
			\If{$\|\mathbf{v}_{i}\| > \left\| \boldsymbol{\varphi}_{f} \right\| \epsilon_{2} $}
			\State $\mathbf{y}_{0}$ $\gets$ $\mathbf{y}_{f}$
			\State Compute $\mathbf{y}_{i}$ recursively using $\eqref{eq:memory_term_creation}$
			\State Compute $\mathbf{\Phi}_{b_{i}}$ and $\mathbf{Y}_{b_{i}}$ using $\eqref{eq:memory_term_creation}$ 
			\State i=i+1
			\EndIf
			\EndIf
			\EndIf
			\caption{To create $\mathbf{\Phi}_{b}$ and $\mathbf{Y}_{b}$ $(\ref{eq:filtered_dynamics_basis})$}
			\label{alg:identification}
		\end{algorithmic}
	\end{algorithm}

	
	\begin{remark}
		Let $\hat{\mathbf{W}}$ denotes the online estimate of the ideal parameter $\mathbf{W}$, and $\tilde{\mathbf{W}} = \hat{\mathbf{W}} - \mathbf{W}$ is the estimation error. The memory term with the proposed algorithm is $\left( \mathbf{Y}_{m} - \hat{\mathbf{W}}^{T} \mathbf{\Phi}_{m} \right)^{T}$, from $\eqref{eq:filtered_dynamics_memory}$ putting $\mathbf{Y}_{m} = \mathbf{W}^{T} \mathbf{\Phi}_{m}$, we obtain that the memory term is $-\tilde{\mathbf{W}}$. Therefore, when this memory term is supplied to an update law $\dot{\hat{\mathbf{W}}}$, it generates an identity matrix as the coefficient matrix of the estimation error dynamics $\dot{\tilde{\mathbf{W}}}$. However, we are interested in extracting the system parameters directly, we are not proposing an update law $\dot{\hat{\mathbf{W}}}$.
	\end{remark}
	
	The estimate of the system parameters $\mathbf{W}$ are stored in the matrix $\mathbf{Y}_{m}$ after processing the filtered system data through \textit{Algorithm} $\ref{alg:identification}$ and post-multiplying by $\mathbf{\Phi}_{b}^{T}$. We propose few transformations of the matrix $\mathbf{Y}_{m}$ to extract the unknown system parameters $\mathbf{A}$, $k_{p}$, and $\boldsymbol{\theta}$.
	
		The following transformations
		\begin{equation}
			\begin{aligned}
				\mathbf{Y}_{m} \begin{bmatrix} \mathbf{I}_{n \times n} & \mathbf{0}_{n \times \left( 1+p \right)} \end{bmatrix}^{T} =& \mathbf{A}, \\
				\mathbf{Y}_{m} \begin{bmatrix} \mathbf{0}_{1 \times n} & 1 & \mathbf{0}_{1 \times p} \end{bmatrix}^{T} =& \mathbf{b} k_{p}, \\
				\mathbf{Y}_{m} \begin{bmatrix} \mathbf{0}_{p \times \left( n+1 \right)} & \mathbf{I}_{p \times p} \end{bmatrix}^{T} =& \mathbf{b} k_{p} \boldsymbol{\theta}^{T},
			\end{aligned}
			\label{eq:transformation}
		\end{equation} generates the ideal values of the system parameters under the FE condition. 
	
	\begin{remark}
		\cite{glushchenko2022exponentially} extracted the ideal values of the system parameters using DREM. A time-varying gain is proposed to negate the influence of the excitation level on the process. Low excitation levels infer high gains, which is never a desirable property. The proposed algorithm does not depend on the excitation level of the regressor vector hence avoiding the use of time-varying gains. Therefore, the designed method has significantly improved properties than the existing methods.
	\end{remark}
	
	\subsection{Proposed Combined Adaptive Control}
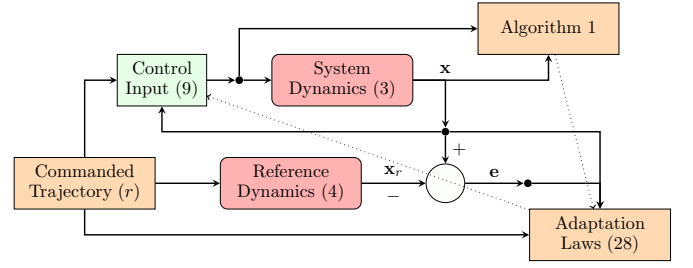
\begin{figure}
		\centering
		\resizebox{1\linewidth}{!}{
			\begin{tikzpicture}[node distance=2cm]
				\node (reference) [process, text width=2.5cm] {Commanded Trajectory $\left(r\right)$ };
				\node (referencedynamics) [startstop, right of=reference, xshift=2cm, text width=2.5cm] {Reference Dynamics $\eqref{eq:reference_dynamics}$};
				\node (add) [summer, right of=referencedynamics, xshift=1cm]{};
				\node (route1) [route, above of=add, yshift=-1cm]{};
				\node (route3) [route, right of=add, xshift=-0.4cm]{};
				\node (control) [io, above of=referencedynamics, xshift=-2.5cm, text width=1.5cm] {Control Input $\eqref{eq:control_law}$};
				\node (route2) [route, right of=control, xshift=-0.5cm]{};
				\node (systemdynamics) [startstop, right of=control, xshift=1.5cm, text width=2.5cm] {System Dynamics $\eqref{eq:system_dynamics}$};
				\node (algorithm) [process, right of=systemdynamics, xshift=2cm, yshift=1cm, text width=2.5cm] {Algorithm $1$};
				\node (adaptationlaws) [process, right of=referencedynamics, xshift=4cm, yshift=-1cm, text width=2.5cm] {Adaptation Laws $\eqref{eq:adaptation_laws_proposed}$};
				
				\draw [arrow] (reference) |- (control);
				\draw [arrow] (control) -- (route2);
				\draw [arrow] (route2) -- (systemdynamics);
				\draw [arrow] (route2) |- (algorithm);
				\draw [arrow] (systemdynamics) -| (algorithm);
				\draw [arrow] (reference) -- (referencedynamics);
				\draw [arrow] (referencedynamics) -- node[anchor=south] {$\mathbf{x}_{r}$} node[anchor=north] {$-$} (add);
				\draw [arrow] (systemdynamics) -| node[anchor=south] {$\mathbf{x}$} (route1);
				\draw [arrow] (route1) -- node[anchor=west] {$+$} (add);
				\draw [arrow] (route1) -| (control);
				\draw [arrow] (reference) |- (adaptationlaws);
				\draw [arrow] (add) -- node [anchor=south] {$\mathbf{e}$} (route3);
				\draw [arrow] (route3) -| (adaptationlaws);
				\draw [arrow] (route1) -| (adaptationlaws);
				\draw [dotted,->] (algorithm) -- (adaptationlaws);
				\draw [dotted,->] (adaptationlaws) -- (control);
			\end{tikzpicture}
		}
		\caption{Block diagram of the closed-loop system.}
		\label{fig:block_diagram}
	\end{figure}

	The ideal values of the system matrices are obtained by performing transformations with the matrix $\mathbf{Y}_{m}$. Let us define a few terms utilizing these transformations that will be used in proposing new adaptation laws. 
	\begin{equation}
		\begin{aligned}
			\mathbf{E}_{1} =& \mathbf{A}_{r} - \mathbf{Y}_{m} \begin{bmatrix} \mathbf{I}_{n \times n} & \mathbf{0}_{n \times \left( 1+p \right)} \end{bmatrix}^{T} \\
			& - \mathbf{Y}_{m} \begin{bmatrix} \mathbf{0}_{1 \times n} & 1 & \mathbf{0}_{1 \times p} \end{bmatrix}^{T} \hat{\mathbf{k}}_{x}^{T}, \\
			\mathbf{E}_{2} =& \mathbf{b}_{r} - \mathbf{Y}_{m} \begin{bmatrix} \mathbf{0}_{1 \times n} & 1 & \mathbf{0}_{1 \times p} \end{bmatrix}^{T} \hat{k}_{r}, \\
			\mathbf{E}_{3} =& \mathbf{Y}_{m} \begin{bmatrix} \mathbf{0}_{p \times \left( n+1 \right)} & \mathbf{I}_{p \times p} \end{bmatrix}^{T} \\
			&- \mathbf{Y}_{m} \begin{bmatrix} \mathbf{0}_{1 \times n} & 1 & \mathbf{0}_{1 \times p} \end{bmatrix}^{T} \hat{\boldsymbol{\theta}}^{T}.
		\end{aligned}
		\label{eq:new_errors}
	\end{equation}
	The proposed adaptation laws for updating the parameter estimates using the tracking error and the above-defined errors are as follows,
	\begin{equation}
		\begin{aligned}
			\dot{\hat{\mathbf{k}}}_{x} =& - \mathbf{x} \mathbf{e}^{T} \mathbf{P} \mathbf{b} k_{p}' + \eta \mathbf{E}_{1}^{T} \mathbf{b} k_{p}', \\
			\dot{\hat{k}}_{r} =& - r \mathbf{e}^{T} \mathbf{P} \mathbf{b} k_{p}' + \eta \mathbf{E}_{2}^{T} \mathbf{b} k_{p}', \\
			\dot{\hat{\boldsymbol{\theta}}} =&\, \boldsymbol{\phi} \mathbf{e}^{T} \mathbf{P} \mathbf{b} k_{p}' + \eta \mathbf{E}_{3}^{T} \mathbf{b} k_{p}',
		\end{aligned}
		\label{eq:adaptation_laws_proposed}
	\end{equation}
	where $\eta = \text{det} \left( \mathbf{\Phi}_{m} \right)$, from  $\eqref{eq:transformation}$, and using the matching condition $\eqref{eq:matching_condition}$, we obtain
	\begin{equation*}
		\mathbf{E}_{1} = - \mathbf{b} k_{p} \tilde{\mathbf{k}}_{x}^{T}, \quad \mathbf{E}_{2} = - \mathbf{b} k_{p} \tilde{k}_{r}, \quad \mathbf{E}_{3} = - \mathbf{b} k_{p} \tilde{\boldsymbol{\theta}}^{T},
	\end{equation*}
	and putting these simplified expressions of the errors in $\eqref{eq:adaptation_laws_proposed}$, we get the following estimation error dynamics
	\begin{equation}
		\begin{aligned}
			\dot{\tilde{\mathbf{k}}}_{x} =& - \mathbf{x} \mathbf{e}^{T} \mathbf{P} \mathbf{b} k_{p}' - \eta \tilde{\mathbf{k}}_{x} \left| k_{p} \right| \mathbf{b}^{T} \mathbf{b}, \\
			\dot{\tilde{k}}_{r} =& - r \mathbf{e}^{T} \mathbf{P} \mathbf{b} k_{p}' - \eta \tilde{k}_{r} \left| k_{p} \right| \mathbf{b}^{T} \mathbf{b}, \\
			\dot{\tilde{\boldsymbol{\theta}}} =&\, \boldsymbol{\phi} \mathbf{e}^{T} \mathbf{P} \mathbf{b} k_{p}' - \eta \tilde{\boldsymbol{\theta}} \left| k_{p} \right| \mathbf{b}^{T} \mathbf{b}.
			\label{eq:estimation_error_dynamics_proposed}
		\end{aligned}
	\end{equation}
	The complete block diagram of the closed-loop system is illustrated in Fig. $\ref{fig:block_diagram}$ for better understanding of the signal flow and the interconnections between the proposed algorithm, adaptation laws, and the control input fed to the system dynamics $\eqref{eq:system_dynamics}$.
	\begin{thm}
		Consider the class of uncertain nonlinear system $\eqref{eq:system_dynamics}$, driven by the control law $\eqref{eq:control_law}$, and the estimate of the controller gains are updated by the adaptation laws $\eqref{eq:adaptation_laws_proposed}$. The regressor vector $\boldsymbol{\varphi}$ satisfies the finite excitation condition after a time interval $t=t_{q}>t_{0}$. The closed-loop system described by $\eqref{eq:tracking_error_dynamics}$ and $\eqref{eq:estimation_error_dynamics_proposed}$ exhibits following properties.
		\begin{enumerate}
			\item The origin of the closed-loop system is uniformly stable $\forall$ $t \geq t_{0}$.
			\item The origin of the closed-loop system is exponentially stable $\forall$ $t > t_{q}$.
			\item The combined error vector satisfies 
			\begin{equation*}
				\left\| \boldsymbol{\chi}(t) \right\| \leq e^{-\kappa \left( t - t_{q} \right)} \alpha \left\| \boldsymbol{\chi}(t_{0}) \right\|, \quad \forall \, t > t_{q},
			\end{equation*}
			where $\alpha \in \mathbb{R}^{+}$, and $\kappa \in \mathbb{R}^{+}$ is the convergence rate, independent of the excitation level of the regressor vector $\boldsymbol{\varphi}$.
		\end{enumerate}
	\end{thm}
	\begin{pf}
		Consider the Lyapunov function $\eqref{eq:lyapunov_function}$, and the time interval when the regressor vector is not satisfying the finite excitation condition, an orthogonal matrix $\mathbf{\Phi}_{b}$ can not be created, inferring $\eta = \text{det} \left( \mathbf{\Phi}_{m} \right) = 0$, hence the time derivative of the Lyapunov function along the error trajectories $\eqref{eq:tracking_error_dynamics}$ and $\eqref{eq:estimation_error_dynamics_proposed}$, and using $\eqref{eq:lyapunov_equation}$ is $\dot{V} \leq -\lambda_{\text{min}} \left( \mathbf{Q} \right) \left\| \mathbf{e} \right\|^{2}$, hence uniform stability follows from the \textit{Proposition} $\ref{prop:proposition_gradient}$. Now, after the time interval $t=t_{q}$, the regressor vector satisfies the finite excitation condition, and we can create an orthogonal matrix $\mathbf{\Phi}_{b}$, and $\mathbf{\Phi}_{m} = \mathbf{I}$, inferring $\eta=1$, hence the time derivative of the Lyapunov function along the error trajectories $\eqref{eq:tracking_error_dynamics}$ and $\eqref{eq:estimation_error_dynamics_proposed}$, and using $\eqref{eq:lyapunov_equation}$, we obtain
		\begin{equation*}
			\dot{V} \leq - \lambda_{\text{min}} \left( \mathbf{Q} \right) \left\| \mathbf{e} \right\|^{2} - 2 \left| k_{p} \right|^{2} \mathbf{b}^{T} \mathbf{b} \left\{ \left\| \tilde{\mathbf{k}}_{x} \right\|^{2} + \tilde{k}_{r}^{2} + \left\| \tilde{\boldsymbol{\theta}} \right\|^{2} \right\},
		\end{equation*}
		in terms of combined error vector $\boldsymbol{\chi}$, we can write
		\begin{equation}
			\dot{V} \leq - \text{min} \left\{ \lambda_{\text{min}} \left( \mathbf{Q} \right), 2 \left| k_{p} \right|^{2} \mathbf{b}^{T} \mathbf{b} \right\} \left\| \boldsymbol{\chi} \right\|^{2},
			\label{eq:V_dot_proposed_1}
		\end{equation}
		the Lyapunov function satisfies, $\text{min} \left\{ \lambda_{\text{min}} \left( \mathbf{P} \right),\left| k_{p} \right| \right\} \left\| \boldsymbol{\chi} \right\|^{2}$ $\leq V \leq$ $\text{max} \left\{ \lambda_{\text{max}} \left( \mathbf{P} \right),\left| k_{p} \right| \right\} \left\| \boldsymbol{\chi} \right\|^{2}$, rewriting $\eqref{eq:V_dot_proposed_1}$,
		\begin{equation}
			\dot{V} \leq -\overline{\kappa} V, \quad \overline{\kappa} = \dfrac{\text{min} \left\{ \lambda_{\text{min}} \left( \mathbf{Q} \right), 2\left| k_{p} \right|^{2} \mathbf{b}^{T} \mathbf{b} \right\}}{\text{max} \left\{ \lambda_{\text{max}} \left( \mathbf{P} \right),\left| k_{p} \right| \right\}},
			\label{eq:V_dot_proposed_2}
		\end{equation}
		where $\overline{\kappa}$ is the convergence rate. The solution of $\eqref{eq:V_dot_proposed_2}$ is
		\begin{equation}
			V\left( t \right) \leq e^{-\overline{\kappa} \left( t-t_{q} \right)} V\left( t_{q} \right), \quad \forall \, t>t_{q}
			\label{eq:V_solution_proposed}
		\end{equation}
		inferring that the origin of the closed-loop system is exponentially stable $\forall \, t>t_{q}$. \par
		For time interval $t\leq t_{q}$, $\dot{V} \leq 0$, inferring $V\left(t\right) \leq V\left(t_{0}\right)$, hence $V\left(t_{q}\right) \leq V\left(t_{0}\right)$, the Lyapunov function satisfies the following, $\text{min} \left\{ \lambda_{\text{min}} \left( \mathbf{P} \right),\left| k_{p} \right| \right\} \left\| \boldsymbol{\chi} \right\|^{2}$ $\leq V \leq$ $\text{max} \left\{ \lambda_{\text{max}} \left( \mathbf{P} \right),\left| k_{p} \right| \right\} \left\| \boldsymbol{\chi} \right\|^{2}$, we can write,
		\begin{equation*}
			\begin{aligned}
				\left\| \boldsymbol{\chi} \left( t \right) \right\|^{2} & \leq \dfrac{1}{\text{min} \left\{ \lambda_{\text{min}} \left( \mathbf{P} \right),\left| k_{p} \right| \right\}} V\left(t\right) \quad \forall \, t \geq t_{0}, \\
				& \leq \dfrac{1}{\text{min} \left\{ \lambda_{\text{min}} \left( \mathbf{P} \right),\left| k_{p} \right| \right\}} V\left(t_{q}\right) \quad \forall \, t > t_{q}, \\
				& \leq \dfrac{e^{-\overline{\kappa} \left( t-t_{q} \right)}}{\text{min} \left\{ \lambda_{\text{min}} \left( \mathbf{P} \right),\left| k_{p} \right| \right\}} V \left( t_{0} \right) \quad \forall \, t > t_{q},
			\end{aligned}
		\end{equation*}
		\begin{equation*}
			\left\| \boldsymbol{\chi} \left( t \right) \right\| \leq e^{-\kappa \left( t - t_{q} \right)} \alpha \left\| \boldsymbol{\chi}(t_{0}) \right\| \quad \forall \, t > t_{q}, 
		\end{equation*}
		where $\kappa = \dfrac{1}{2}\overline{\kappa}$, $\alpha = \sqrt{\dfrac{\text{max} \left\{ \lambda_{\text{max}} \left( \mathbf{P} \right),\left| k_{p} \right| \right\}}{\text{min} \left\{ \lambda_{\text{min}} \left( \mathbf{P} \right),\left| k_{p} \right| \right\}}}$, $\kappa$ is dependent on gains $\mathbf{Q}$ and $\mathbf{P}$, and the system parameters $\mathbf{b}$ and absolute of $k_{p}$.  \hfill $\blacksquare$
	\end{pf}
	
	\begin{remark}
		In the proposed \textit{Algorithm} $\ref{alg:identification}$, we require initialization of two variables $\epsilon_{1}$ and $\epsilon_{2}$. The choice of $\epsilon_{1}$ depends on the operating region of the system and the other variable $\epsilon_{2}$ is the residual in the MGS process. It is to be noted that the choice of both variables does not affect the property of independence of the convergence rate on the excitation level of the regressor vector under the FE condition.
	\end{remark}
	
	\begin{remark}
		We have chosen unity gains in the adaptation laws $\eqref{eq:adaptation_laws_proposed}$ that can be selected as other valid values and the stability analysis is trivial with the appearance of the chosen adaptation gains in the convergence rate $\kappa$.  
	\end{remark}
	

	\section{Simulation Results}
	\label{section:Simulation_Results}
\begin{figure}
		\centering
		\includegraphics[width=1\linewidth]{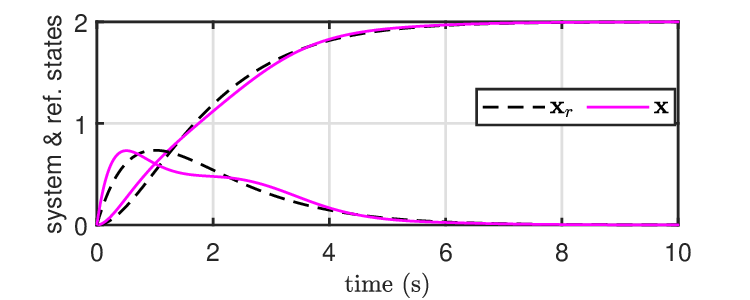}
		\caption{Tracking of reference state $\mathbf{x}_{r}$ by system state $\mathbf{x}$.}
		\label{fig:states}
	\end{figure}

\begin{figure}
		\centering
		\includegraphics[width=1\linewidth]{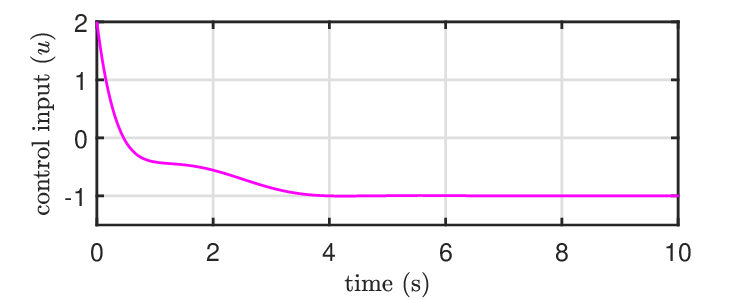}
		\caption{Control input $u$.}
		\label{fig:control_input}
	\end{figure}

\begin{figure}
		\centering
		\includegraphics[width=1\linewidth]{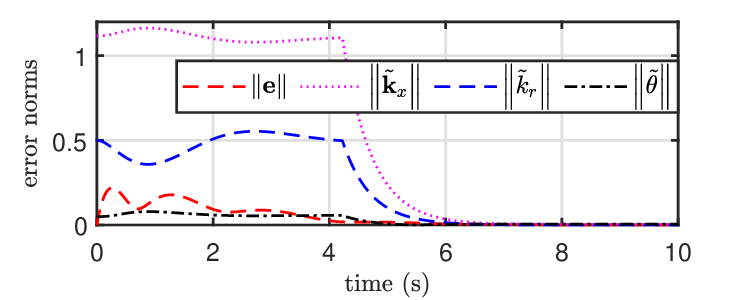}
		\caption{Error norms.}
		\label{fig:error_norms}
	\end{figure}

\begin{figure}
		\centering
		\includegraphics[width=1\linewidth]{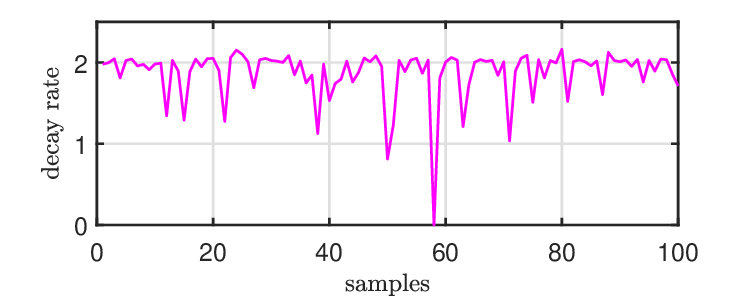}
		\caption{Decay rate for $100$ samples.}
		\label{fig:decay_rate}
	\end{figure}

\begin{figure}
		\centering
		\includegraphics[width=1\linewidth]{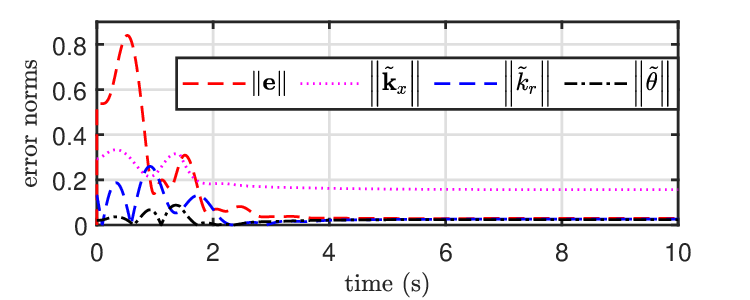}
		\caption{Error norms for $58^{\text{th}}$ sample.}
		\label{fig:error_norms_58}
	\end{figure}

	We consider the following uncertain nonlinear system for validation of the proposed results,
	\begin{equation*}
		\begin{bmatrix} \dot{x}_{1} \\ \dot{x}_{2}   \end{bmatrix} = \begin{bmatrix} 0 & 1 \\ 1 & 0 \end{bmatrix} \begin{bmatrix} x_{1} \\ x_{2}   \end{bmatrix} + \begin{bmatrix} 0 \\ 1 \end{bmatrix}  \left( 2 \right) \left\{ u + \left( -0.1 x_{2}^{2} \right) \right\}.
	\end{equation*}
	The reference dynamics is as follows,
	\begin{equation*}
		\begin{bmatrix} \dot{x}_{r_{1}} \\ \dot{x}_{r_{2}}   \end{bmatrix} = \begin{bmatrix} 0 & 1 \\ -1 & -2 \end{bmatrix} \begin{bmatrix} x_{r_{1}} \\ x_{r_{2}}   \end{bmatrix} + \begin{bmatrix} 0 \\ 1 \end{bmatrix} r, \quad \mathbf{Q} = \begin{bmatrix} 1 & 0 \\ 0 & 1  \end{bmatrix},
	\end{equation*}
	where $\mathbf{Q}$ is required for $\mathbf{P}$ $\eqref{eq:lyapunov_equation}$. On comparing with linear in parameter form $\eqref{eq:system_lip}$, we obtain $\mathbf{W} = \begin{bmatrix} 1 & 0 & 2 & -0.2 \end{bmatrix}^{T}$, and  $\boldsymbol{\varphi} = \begin{bmatrix} x_{1} & x_{2} & u & x_{2}^{2} \end{bmatrix}^{T}$. The algorithm parameters are chosen as $\epsilon_{1}=1$, $\epsilon_{2}=0.01$, and filter frequency $f=1$. The commanded signal is $r(t)=2$, and we consider a $50\%$ error in the initial estimate of the unknown parameters. The initial conditions of the system and reference system states are chosen to be zero. Fig. $\ref{fig:states}$ shows the time evolution of reference state tracking by the system state and the required control input is shown in Fig. $\ref{fig:control_input}$. Fig. $\ref{fig:error_norms}$ shows all the error norms and it can be observed that they decay to the origin exponentially fast after the FE condition is satisfied. Further, we have run simulations for $100$ samples in which the commanded signal, error percentage in the initial estimate of the system parameters, and system states' initial values are varied between $\left[2,6\right]$, $\left[20\%,80\%\right]$, $\left[0,1\right]$, and $\left[-0.1,0.1\right]$, respectively. The decay rate is computed for each sample by calculating the time difference from the satisfaction of the FE condition to the time $\left\| \boldsymbol{\chi} \left(t_{\text{i}}\right) \right\| \leq 2\% \left\| \begin{bmatrix} \mathbf{x}_{r}^{T} & \mathbf{k}_{x}^{T} & k_{r} & \theta \end{bmatrix}^{T} \right\|$ satisfies, where $t_{\text{i}}$ is a time instant. The convergence rate $\overline{\kappa}$ $\eqref{eq:V_dot_proposed_2}$ for the given system is $\overline{\kappa}=0.5$ inferring $\kappa=0.25$. Fig. $\ref{fig:decay_rate}$, shows the decay rate for the uniformly random $100$ samples, and the convergence rate for all the samples is well within the limits except for one sample. We identified the randomly generated commanded signal, percentage uncertainty in the initial estimate of the parameters, and the initial state values for the identified $58^{\text{th}}$ sample. The error norms for the $58^{\text{th}}$ sample is shown in Fig. $\ref{fig:error_norms_58}$. We observe that the combined error norm is not within the $2\%$ limit causing no computation of the decay rate. This may be attributed to erroneous collection of the data in $\mathbf{Y}_{m}$ reflecting the error norm to be not decaying below the $2\%$ limit. Therefore, collection of several data points would be beneficial in improving the accuracy.

	\section{Conclusion}
	\label{section:Conclusion}
	The developed exponentially stable combined adaptive control for systems with a known sign of the control effectiveness vector may be extended to multi-input-multi-output systems. The proposed algorithm may be applied to solve the tracking/regulation problems for the systems with slowly time-varying parameters. The effectiveness of the proposed method may be validated in designing fault-tolerant adaptive control systems.

	
	\bibliography{ifacconf}             

\end{document}